\documentstyle[prl,multicol,aps]{revtex}
\begin{document}

\draft

\title
{ Exact results for a Kondo impurity with electronic correlations }

\author
{You-Quan Li}

\address
{
Institut de Physique Th\'{e}orique, 
\'{E}cole Polytechnique F\'{e}d\'{e}rale de Lausanne, 
CH-1015 Lausanne \\
and Zhejiang Institute of Modern Physics,
Zhejiang University, Hangzhou 310027, P. R. China 
}

\author
{Pierre-Antoine Bares}

\address
{
Institut de Physique Th\'{e}orique, 
\'{E}cole Polytechnique F\'{e}d\'{e}rale de Lausanne, 
CH-1015 Lausanne \\
}

\date{ Received July 7, 1997 }

\maketitle

\begin{abstract}

An one dimensional  model of a Kondo impurity with arbitrary spin in the
presence of electronic correlations is solved exactly. The Bethe ansatz 
equations are obtained for the case of quadratic dispersion. 
The ground state is studied in the thermodynamic limit, and  
the magnetic susceptibility at zero
temperature is calculated.
\end{abstract}

\pacs{PACS number(s): 75.20.Hr; 72.15.Qm }

\begin{multicols}{2}

It is known that the study of exact solutions is 
useful to the understanding of non-perturbative effects in the 
strongly correlated electronic systems. 
The exact solution of the one dimensional 
Kondo model with linearized dispersion and no correlations between
electrons 
was found in \cite{Andrei,Wiegmann}.
In this paper, we study the problem of a Kondo impurity with arbitrary
spin in the presence of the electronic correlations.
We show that the model is exactly solvable at some value of 
electron-impurity coupling.
Using Bethe  ansatz, we solve the  
secular equations for the spectrum with periodic boundary condition.
The ground state as well as the magnetic susceptibility
at zero temperature are calculated.

We focus on a  system described  by the Hamiltonian 
\begin{eqnarray}
H_0 = \sum_{\bf k }\varepsilon({\bf k}) C^{\ast}_{ {\bf k} a} C_{ {\bf
k} a} 
        \hspace{36mm}\nonumber\\
      + \sum_{ 
              {\bf k}_1 , {\bf k}_2 , {\bf k}_3 ,{\bf k}_4 
              }
        u \delta ( {\bf k}_1 +{\bf k}_2 , {\bf k}_3 +{\bf k}_4 ) 
         C^{\ast}_{ {\bf k}_4 a} C^{\ast}_{ {\bf k}_3 a} 
         C_{ {\bf k}_2 a} C_{ {\bf k}_1 a},\nonumber  
\end{eqnarray}      
where $C_{ {\bf k} a }$ annihilates an electron with 
momentum ${\bf k}$ and spin component $a$. The electrons are coupled by
both spin and  charge interactions to a localized impurity,
\[ 
H_I =  J\Psi^{\ast}_a (0) {\bf S}_{ab}\Psi_{b}(0) \cdot {\bf S}^0 
     + V\Psi^{\ast}_a (0) \Psi_a (0),
\]
where the field $\Psi_a$ is the Fourier transform of $ C_{ {\bf k} a}$,
${\bf S}^0 $ is the spin of the impurity and  
${\bf S}$ is the spin  of the electrons in the  band.
If we restrict our attention to the 
low-energy regime, namely, ${\bf k}$ close to the Fermi surface, 
we obtain the (linear dispersion) one-dimensional Kondo Hamiltonian  
with a $\delta$-function 
interaction between electrons.
The coupling  between the electrons does not 
contribute to the scattering matrix (S-matrix) in the case of 
linear dispersion.
However, once correlations are introduced, the scattering matrix
is determined uniquely from the Schr\"{o}dinger equation and 
the anti-symmetry of the fermionic wave function.

We consider the case of quadratic dispersion, i.e.,
$\varepsilon(k)= k^{2}/2 $ (in units of $\hbar$ and of the electron
mass).
We show that the present model can be solved
exactly by means of Bethe ansatz with periodic boundary conditions.
In order to have a  periodic Hamiltonian  in the Hilbert space of
$N$ electrons, we define the following anticommutation relations for the
fields,
\[
\{\Psi(x), \, \Psi^{\dagger} (y) \}
 = \sum_{ n \in Z \!\!\! Z} \delta(x - y - nL).
\]
Then the Hamiltonian in the Hilbert space takes the form
\begin{eqnarray}
H = - \displaystyle\frac{1}{2}\sum_{j=1}^{N}
       \frac{\partial^{2} }{\partial x_{j}^{2} } 
  + \sum_{\stackrel{i<j}{n\in Z\!\!\!Z } } u \delta(x_i - x_j - nL)
    \nonumber\\ 
+ \sum_{\stackrel{j=1}{n\in Z\!\!\!Z} }^{N} 
    (J {\bf S}^j \cdot {\bf S}^0  + V )\delta(x_j - nL).
 \label{eq:hamiltonian}
\end{eqnarray}
The Hamiltonian operator (\ref{eq:hamiltonian}) is invariant under
arbitrary permutation. Although it does not have 
translational symmetry due to the
presence of the impurity, (\ref{eq:hamiltonian}) 
is invariant under a translation
of finite distance  $L$. 
Let $\{ {\sf l \!\! P}_{\alpha} \}$ be the set of hyperplanes 
${\sf l \!\! P}_{\alpha}:= \{ x | (\alpha | x ) = 0 \}$, 
where the notations 
$\, x:=(x_1, x_2, \cdots, x_N)$, $\alpha= e_i - e_j , \, e_i$, and
$(\alpha | x ) := \sum_{i=1}^{N} (\alpha)_i x_i $ are used. 
In the domain ${\sf l \!\! R}^N \setminus \{ {\sf l \!\! P}_{\alpha}
\}$,
eq.(\ref{eq:hamiltonian}) becomes an $N$-dimensional Laplace operator,
so 
$N$-dimensional plane waves solve the Schr\"{o}dinger equation. 
As those hyperplanes partition the domain into finitely many regions,
the wave function must be a piecewise continuous function.
These regions can be labeled by both an element of 
the permutation group $S_N$ and a number indicating the ordering 
of the impurity and the electrons, namely, 
${\cal C}(Q^{(i)}):=\{ x | x_{Q1} <\cdots < x_{Qi}< 0 <\cdots < x_{QN}
\}$,
here $i = 0,1,2, \cdots, N$ and $Q^{(0)}$ means 
$ 0 < x_{Q1} <\cdots < x_{QN} $.  
In  these notations we can write the wave function in the Bethe
ansatz form,
\begin{equation}
\psi_a (x)=\sum_{P \in S_N } 
           A_a (P; \, Q^{(i)} ) 
           e^{i(Pk|Qx)},
\,\, for  \,  x \in {\cal C}( Q^{(i)} ),
\label{eq:BAW}
\end{equation}
where $ a:=(a_{Q1}, a_{Q2}, \cdots, a_{QN})$, $a_j$ 
denotes the spin component of the $j$th particle; 
$Pk$ represents the image of a given $k:=(k_1, k_2, \cdots, k_N )$ by 
a mapping $P \in S_N $; $(Pk | Qx) = \sum_{j=1}^{N} (Pk)_j (Qx)_j$. 

The wave function for the electrons must be anti-symmetric 
under any permutation
of both the coordinates and the spin components of the electrons. 
Since any element
of the permutation group $S_N$ can be expressed 
as a product of the neighboring 
interchanges,
$ 
\sigma^j : ( \cdots, z_j, z_{j+1}, \cdots ) 
             \mapsto ( \cdots, z_{j+1}, z_j, \cdots )
$,
where $z_j$ denotes either $x_j, a_j$ or their image by a mapping of
$S_N$.
The requirement of anti-symmetry is  
$
(\sigma^j \psi )_a (x) = - \psi_a (x).
$
Using the identity 
$
( Pk | \sigma^j x ) = (\sigma^j Pk |  x )
$, 
we obtain             
\begin{equation}
A(P; \, \sigma^j Q^{(i)} ) = - {\cal P}^{Qj,Q(j+1)} A(\sigma^j P; \,
Q^{(i)} ),
\label{eq:Antisymmetry}
\end{equation}
where the spin labels are omitted and
${\cal P}^{Qj,Q(j+1)}$ is the spinor representation of the permutation
$\sigma^j $. Explicitly, we have 
${\cal P}^{Qj,Q(j+1)}= ( 1 + {\bf S}^{Qj}\cdot{\bf S}^{0} )/2$. 
The $\delta$-function terms in the Hamiltonian (\ref{eq:hamiltonian})
contribute boundary conditions at hyperplanes 
$\{ {\sf l \!\! P}_{\alpha} \}$\cite{Sutherland,LiMa1}, 
i.e., the discontinuity
of the derivatives of the wave function along the normal of the
hyperplanes:
\begin{eqnarray}
\lim_{\epsilon\rightarrow 0^+}
  \left\{
       \alpha\cdot\nabla 
         \left[ 
               \psi_a (x_{(\alpha)} + \epsilon\alpha)
             - \psi_a (x_{(\alpha)} - \epsilon\alpha)
         \right] \right.      \nonumber \\  \left.
  - U\left[ 
           \psi_a (x_{(\alpha)} + \epsilon\alpha)
           + \psi_a (x_{(\alpha)} - \epsilon\alpha)
     \right]     
 \right\} = 0,  
\label{eq:BC}
\end{eqnarray}
where $\nabla := \sum_{j=1}^{N} e_j (\partial /\partial x_j )$,
$x_{(\alpha)} \in {\sf l \!\! P}_{\alpha}$, 
$U=u$ for $\alpha = e_i - e_j $ and 
$ U= J {\bf S}^j \cdot {\bf S}^0 + V $ for $\alpha = e_i $.

As the regions ${\cal C}(Q^{(i)} )$ and 
${\cal C}(\sigma^j Q^{(i)} ) \, (j \neq i) $ 
are adjacent to each other, and so are ${\cal C}(Q^{(i)} )$ and 
${\cal C}(Q^{(i-1)}) $, 
the respective boundary conditions (\ref{eq:BC}) 
relate the wave functions on neighboring regions. 
After writing out the boundary conditions (\ref{eq:BC})
and using (\ref{eq:BAW}), we find that: 
$
A(P; \sigma^j Q^{(i)}) = S^{Qj,Q(j+1)}[ (Pk)_{j}, \, (Pk)_{j+1} ]   
                      A(P; Q^{(i)})
$ for $j \neq i$, and 
$
A(P; Q^{(i-1)}) = S^{Qi, 0}[ (Pk)_i ] A(P; Q^{(i)})
$.
The S-matrix $S^{Qj, 0}$ of the electron-impurity depends on 
the coupling constants
$J$ and $V$, and the S-matrix of electrons takes the same form as  
that of the $\delta$-function fermion gas\cite{Gaudin,Yang}. 

Both the electron-electron and electron-impurity S-matrices  
relate the coefficients $A$ between distinct regions in the 
configuration space of $N$ electrons.
Using (\ref{eq:Antisymmetry}) and the S-matrices, 
we find that the coefficients $A$
in any region are determined up to an overall factor 
by the $\check{S}$-matrices,
$\check{S}^{Qi,Q(j+1)}:= - {\cal P}^{Qj,Q(j+1)} S^{Qj,Q(j+1)}$,\\ i.e.,
\,\,
$A(\sigma^i P; Q^{(i)}) \,=\,
 S^{Q(i+1), 0}\,\check{S}^{Qi,Q(i+1)}\,[S^{Q(i+1), 0}]^{-1} 
\\ A(P; Q^{(i)})$, and 
$A(\sigma^j P; Q^{(i)}) = \check{S}^{Qj,Q(j+1)} A(P; Q^{(i)}) $ 
for $j\neq i$.

The Yang-Baxter equations arising from  both
$A(\sigma^j \sigma^{j+1} \sigma^j P; Q^{(i)} ) \,=\,
A(\sigma^{j+1} \sigma^j \sigma^{j+1} P; Q^{(i)} )$ and
$A( P; \sigma^j \sigma^{j+1} \sigma^j Q^{(i)} ) \,= \,
A( P; \sigma^{j+1} \sigma^j \sigma^{j+1} Q^{(i)} )$
for $i \neq j, j+1$ 
are fulfilled  identically. 
However, the others for $ i= j, j+1$, namely,
$S^{Qi,Q(i+1)}S^{Qi,0}S^{Q(i+1),0}= S^{Q(i+1),0}S^{Qi,0}S^{Qi, Q(i+1)}$
are satisfied only when
$J=0$ and $V$ arbitrary, or
$J=-2u$ and  $V= -(s+1)u$ or $su$ for an impurity of spin-$s$. 
The former corresponds to a non-magnetic impurity\cite{LiMa1}.
In the following we will focus on the later case. 
The electron-impurity S-matrix reads
\begin{equation}
S^{Qi,0} = e^{-i\theta [ (Pk)_{i} ] }
           \frac{
                 i(Pk)_{i} -  u {\cal P }^{Qi, 0}               
                }{
                 i(Pk)_{i} -  u (s + \displaystyle\frac{1}{2} )
                 }.                
\label{eq:Smatrix}
\end{equation}
where $ {\cal P}^{Qi,0} = (1 + {\bf S}^{Qi}\cdot {\bf S}^{0} )/2 $,
$ \theta(x)= 2\tan^{-1}( x /u(s+1/2) ) + \pi $ for $V=-(s+1)u$ and
$ \theta(x)= 0  $ for $V= su$. 
These two cases correspond respectively to the anti-parallel and 
parallel spin-coupling between
electron and impurity.

Now we proceed to determine the secular equations for the spectrum by 
considering periodic boundary condition. If $x$ is a point in the 
region ${\cal C}(Q^{(i)})$, then
$ x'=(x_1, \cdots, x_{Q1}+ L, \cdots, x_N )$ is a point in the region 
${\cal C}(\gamma Q^{(i-1)})$ with
$\gamma = \sigma^{N-1}\sigma^{N-2}\cdots\sigma^{2}\sigma^{1}$.
So the periodic boundary condition imposes a relation between the
wave functions defined on ${\cal C}(Q^{(i)})$ and
${\cal C}(\gamma Q^{(i-1)})$. Writing out this relation
in terms of (\ref{eq:BAW}), we find that the periodic boundary
conditions are guaranteed provided that 
$A(P ; \gamma Q^{(i-1)}) e^{i(Pk)_{1}L }= A(P ; Q^{(i)})$.
After applying  the S-matrices successively, we obtain an eigenvalue 
equation in the spinor space:
\begin{eqnarray}
S^{Q1,QN}\cdots S^{Q1,Q(i+1)}S^{Q1,0}S^{Q1,Qi}\cdots
S^{Q1,Q2}A(P ; Q^{(i)}) \nonumber \\
= e^{-i(Pk)_{1} L} A(P ; Q^{(i)}). \hspace{20mm}
\label{eq:PBC}
\end{eqnarray}

Applying the quantum inverse scattering method (see \cite{Faddeev}),
we find the  Bethe Ansatz equations:
\begin{eqnarray}
e^{-ik_j L }= e^{-i\theta( k_j )}\prod_{\nu=1}^{M}
              \frac{\lambda_{\nu} - k_j + iu/2}
                   {\lambda_{\nu} - k_j - iu/2},
     \hspace{16mm}\nonumber \\
 - \prod_{\nu=1}^{M}\frac{\lambda_{\nu} - \lambda_{\mu} + iu}
                         {\lambda_{\nu} - \lambda_{\mu} - iu}
 = \frac{ \lambda_{\mu} - ius}
        { \lambda_{\mu} + ius}
   \prod_{l=1}^{N}\frac{\lambda_{\mu} - k_l - iu/2}
                       {\lambda_{\mu} - k_l + iu/2}.
\label{eq:BAE}
\end{eqnarray}  
Taking the logarithm of (\ref{eq:BAE}), we obtain a set of
coupled transcendental equations,
\begin{eqnarray}
k_j = (\frac{2\pi}{L})I_j 
   + \frac{1}{L}\left[\sum_{\nu=1}^{M}
                     \Theta_{\frac{1}{2} }(\lambda_{\nu}- k_j)
                   + \eta\Theta_{s + \frac{1}{2} }( k_j )
                \right]                              
\nonumber \\
\sum_{\nu =1}^{M}\Theta_{1}(\lambda_{\mu} - \lambda_{\nu} )
= - 2\pi J_{\mu} 
  + \sum_{l=1}^{N}\Theta_{\frac{1}{2} }(\lambda_{\mu} - k_l )
  + \Theta_{s}
  (\lambda_{\mu} ),    
\label{eq:Secular}
\end{eqnarray}
where $\Theta_{\beta}(x) := 2\tan^{-1} (x/\beta u)$,
$-\pi < \tan^{-1}(x) \leq \pi $; $\eta = 1$
for $V=-(s+1)u$ and $\eta = 0$ for $V=su$. The spin quantum number
$J_{\mu}$
takes integer values or half-integer values according to whether  
$N-M$ is even or odd; The charge quantum number $I_j $ takes integer
values or half-integer
values according to whether $M+1$ (or $M$) is even or odd for
$V=-(s+1)u$ 
(or for $V=su$). The last terms on both right hand sides of 
(\ref{eq:Secular}) represent the contributions of the impurity. 
The phase shift due to the impurity affects the quantum numbers
$I_j$ and $J_\mu$.

We consider the thermodynamic limit. 
The ground state of the present model is a Fermi
sea described by
$\rho_{0}(k)$ and $\sigma_{0}(\lambda)$, where
$\rho_{0}(k)$ is the distribution function of charge with momentum $k$
and $\sigma_{0}(\lambda)$ that of down spins with respect to the
rapidity 
$\lambda$. 
The subscript zero is used for the case of zero magnetic field. 
The distributions of the roots satisfy the following 
coupled integral equations, 
\begin{eqnarray}
  \rho_{0}(k) & = &\frac{1}{ 2\pi }- \frac{\eta }{ L } K_{s+\frac{1}{2}
}(k) 
        + \int^{B}_{-B} d\lambda'\sigma_{0}(\lambda')
           K_{\frac{1}{2} }(k - \lambda'),
  \nonumber \\
  \sigma_{0}(\lambda) 
    &= &  \frac{1}{L } K_{s}(\lambda) 
         - \int^{B}_{-B} d\lambda'\sigma_{0} (\lambda')
           K_{1}(\lambda - \lambda')  \nonumber \\
    &\,& + \int^{D}_{-D} dk'\rho_{0}(k')
           K_{\frac{1}{2} }(\lambda - k'),
 \label{eq:density}
\end{eqnarray}
where $ K_{\beta}(x) := \beta u /( \beta^2 u^2 + x^2 )\pi$.
The $B$ and $D$ are determined from the conditions 
\begin{eqnarray}
\int^{B}_{-B} d\lambda'\sigma_{0}(\lambda') = \frac{M}{L}, \nonumber \\
\int^{D}_{-D} dk'\rho_{0}(k') = \frac{N}{L}.\nonumber
\end{eqnarray} 

The energy of the ground state can be calculated once $\rho_{0}(k)$ 
is known. In the absence of magnetic field $B= \infty$,
we can write (\ref{eq:density}) in a closed form by
Fourier transform,
\begin{eqnarray}
2\pi\rho_{0}(k) = 1 + (\frac{4\pi}{u}) 
           \int_{-D_0}^{D_0}d k' \rho_{0}(k') R_{\frac{1}{2}}
                 \left( \frac{2k - 2k'}{u}
                  \right)   \nonumber \\
         + \frac{1}{L} (\frac{4\pi}{u}) R_s (\frac{2k}{u})
         + \eta\frac{2}{L} K_{s+\frac{1}{2} }(k), 
\end{eqnarray}
where 
\begin{eqnarray}
R_s (x) & = &\frac{1}{4\pi} \int^{\infty}_{-\infty}
                \frac{ e^{-(s-\frac{1}{2} )|y| } }
                     { 1 + e^{|y|} }
                e^{- \frac{ixy}{2} } dy \nonumber \\
        & = & \frac{1}{\pi}\sum^{\infty}_{l=1}
               (-1)^{l-1} \frac{ 2(l + s - \frac{1}{2} ) }
                               { 4 (l + s - \frac{1}{2} )^2 + x^2 }.
\nonumber
\end{eqnarray}                
In the strong coupling limit $ u \gg 1 $, we  obtain 
\begin{equation}
2\pi\rho_{str}= 1 + (\frac{4\pi}{u} )
                \left[ \frac{N}{L} \frac{\ln 2 }{2\pi} 
                 +\frac{1}{L}(\frac{\eta}{2\pi(s+\frac{1}{2} ) }
                  + W_s )
                \right],
\label{eq:Strong}
\end{equation}
where $ W_s = (-1)^s [ 1/4 + \sum^{s}_{l=1} (-1)^l /(2l -1)\pi ]$
for  impurity with {\it s = integer} and $ W_s = (-1)^{s-\frac{1}{2} } 
[\ln2/2\pi + \sum^{s-1/2}_{l=1} (-1)^l /2l \pi ] $
for  impurity with {\it s = half-integer}. 
From the explicit form (\ref{eq:Strong})
we can completely determine $D$, and the energy density of 
the ground state is                                   
\begin{equation}
\frac{E}{L} = \frac{1}{24}\left( 2\pi\frac{N}{L} \right)^3
2\pi\rho_{str}.               
\label{eq:energy}
\end{equation}

In the presence of a magnetic field, we add to the Hamiltonian 
(\ref{eq:hamiltonian}) a Zeemann term,
$ -h \sum_{j=0}^{N} S^j_z $.
We also add  a term arising from the nonvanishing chemical potential.
For $h \neq 0$, $B $ depends on the
magnetic field. It is convenient to 
introduce the dressed energies\cite{FKorepin} of charge and spin, 
$\varepsilon_{\rho}(k)$,  $\varepsilon_{\sigma}(\lambda)$, which are
defined by
\begin{eqnarray}
\frac{E}{L} = \int_{-D}^{D}dk \rho^{(0)}(k)\varepsilon_{\rho}(k)
             +\int_{-B}^{B}d \lambda
             \sigma^{(0)}(\lambda)\varepsilon_{\sigma}(\lambda) 
             - h \frac{s}{L},
\label{eq:dressdef}
\end{eqnarray}
where
$\rho^{(0)} = 1/2\pi - \eta K_{ s+\frac{1}{2} }(k)/L$ and
$\sigma^{(0)} (\lambda) = K_{s}(k)/L$. 
The equations (\ref{eq:density}) of Fredholm-type\cite{Morse} 
for $\rho(k)$ and $\sigma(\lambda)$  are
equivalent to the following integral equations for the dressed energies,
\begin{eqnarray}
\varepsilon_{\rho}(k)
       &=& \varepsilon_{\rho}^{(o)}(k)
           +\int_{-B}^{B}d \lambda \varepsilon_{\sigma}(\lambda')
               K_{ \frac{1}{2} }(\lambda' -k) \nonumber \\
\varepsilon_{\sigma}(\lambda)
       &=& \varepsilon_{\sigma}^{(o)}(\lambda)
           +\int_{-D}^{D}dk' \varepsilon_{\rho}(k)
            K_{ \frac{1}{2} }(\lambda' -k) \nonumber \\
       &\,& - \int_{-B}^{B}d \lambda \varepsilon_{\sigma}(\lambda')
                   K_{1}(\lambda' - \lambda), 
\label{eq:dresseqs}
\end{eqnarray}
where 
$\varepsilon^{(o)}_{\rho,\sigma}$ are the bare energies, namely
$\varepsilon^{(o)}_{\rho} = k^2 /2 + \mu - h/2$,
$\varepsilon^{(o)}_{\sigma} = h$.
The solutions of (\ref{eq:dresseqs}) define the pseudo-energy bands.
The filling of all states with 
$\varepsilon_{\rho}(k) < 0 $ and $\varepsilon_{\sigma}(\lambda)<0$ 
corresponds the ground state. The Fermi levels of both
$k$-sea and $\lambda$-sea are determined by the conditions
$\varepsilon_{\rho}(D) = 0 $ and $\varepsilon_{\sigma}(B) = 0$.

Once the distribution $\sigma(\lambda)$ is solved, 
the magnetization per unit length follows as,
\begin{equation}
{\cal M} = \int_{B}^{\infty} d\lambda \sigma_{B}(\lambda) 
           + \frac{1}{L}(s-\frac{1}{2}).
\label{eq:Mag}
\end{equation}
Clearly, ${\cal M}$ is not exactly zero when 
$B = \infty $ and $ s \neq 1/2$ due to the magnetization of the
impurity. 
We will focus on the case
of the weak magnetic field. Then the $\sigma(\lambda)$
for large $\lambda$ mainly contributes  to the magnetization. 
With the assumption $\lambda \gg D$, the second equation
of (\ref{eq:density}) becomes \cite{YYang}
\begin{equation}
\sigma_{B} (\lambda) 
+ \int_{|\lambda'|>B} d \lambda'
  {\cal K }(\lambda -\lambda')\sigma_{B} (\lambda')
= \sigma_{\infty} (\lambda).
\label{eq:preWH}
\end{equation}
where a new kernel ${\cal K}$ and a distribution function 
$\sigma_{\infty}(\lambda)$ have been introduced. They satisfy the
following
equations
\begin{eqnarray}
{\cal K}(\lambda -\lambda')
&=& K_{1}(\lambda -\lambda')
 + \int_{-\infty}^{\infty}d \lambda''K_{1}(\lambda -\lambda'')
                {\cal K}(\lambda'' -\lambda') \nonumber \\
\sigma_{\infty}(\lambda)
&=& \sigma^{(0)} (\lambda)
  + \int_{-\infty}^{\infty}d \lambda'K_{1}(\lambda -\lambda')
       \sigma_{\infty}(\lambda').
\label{eq:inftysigma}
\end{eqnarray}
Shifting the origin, by introducing 
$\sigma(\lambda) = \sigma_{B}(\lambda + B )$
and neglecting the higher-order terms, we obtain the following
Wiener-Hopf
integral equation for $\sigma(\lambda)$.
\begin{equation}
\sigma (\lambda) 
+ \int_{0}^{\infty} d \lambda'
  {\cal K }(\lambda -\lambda')\sigma (\lambda')
= \sigma_{\infty} (\lambda + B).
\label{eq:WH}
\end{equation}
Equation (\ref{eq:WH}) can be solved by the 
one-side Fourier transform\cite{Morse},
and reads
\begin{equation}
 (\sqrt{2\pi}\tilde{\cal K}(\omega) + 1)\tilde{\sigma}^{+}(\omega) 
 + \tilde{\sigma}^{-}(\omega)
 = \tilde{\sigma}_{\infty}(\omega) e^{-i\omega B}
\label{eq:WH-Fourier}
\end{equation}
with
\begin{equation}
(\sqrt{2\pi}\tilde{\cal K}(\omega) + 1)
= \frac{ e^{\frac{u}{2}|\omega|} }
       { 2 \cosh(\frac{u}{2}\omega ) },
\label{eq:K-Fourier}
\end{equation}
where $\sigma^{+}(\omega)$ and $\sigma^{-}(\omega)$ 
are the analytical parts
of the Fourier transform $\sigma(\omega)$ 
on the upper and lower $\omega$-plane,
respectively. The kernel (\ref{eq:K-Fourier}) is regular in the strip
$-\frac{\pi}{u} < {\rm Im} \omega < \frac{\pi}{u}$, 
and is factorized into 
$\tilde{\cal K}^{+}(\omega)$ and $\tilde{\cal K}^{-}(\omega)$ such that
\begin{equation}
(\sqrt{2\pi}\tilde{\cal K}(\omega) + 1)
=\tilde{\cal K}^{+}(\omega)\left[ \tilde{\cal K}^{-}(\omega)
\right]^{-1},
\nonumber
\end{equation}
where 
$\tilde{\cal K}^{+} = (-iz)^{iz} e^{-iz}\Gamma (1/2 - iz)/\sqrt{2\pi}$
with $ z =\omega u/2\pi$ and 
$[ \tilde{\cal K}^{-}(\omega)]^{-1} = \tilde{\cal K}^{+}(-\omega)$.
$\tilde{\cal K}^{+}(\omega)$ and
$\tilde{\cal K}^{-}(\omega)$ are regular and free of zeros respectively
in the upper and lower half-planes. After seperating 
$\tilde{\cal K}^{-}(\omega) 
\tilde{\sigma}_{\infty}(\omega) e^{-i\omega B }$
into a sum of two parts
$q_{+}(\omega) + q_{-}(\omega)$, which are analytical in upper and 
lower half-planes  respectively, 
we obtain the solution of (\ref{eq:WH-Fourier})
$\tilde{\sigma}^{+}(\omega) = q_{+}(\omega)/\tilde{\cal K}^{+}(\omega)$.
The magnetization at zero temperature is 
\[
{\cal M} = \tilde{\sigma}^{+}(0) + \frac{1}{L} (s- \frac{1}{2} ).
\]
We infer the magnetic susceptibility as 
\begin{equation}
\chi = \frac{1}{L} ( \frac{1}{\sqrt{2}u} \frac{\partial B }{\partial h}
)
  \sum_{n=0}^{\infty} \frac{ (-1)^n }{n!}f_{s}(n)
e^{-(2n+1)\frac{\pi}{u}B}       
\label{eq:susceptibility},
\end{equation}               
where 
$f_{s}(n) = \exp \{ 
-i( n + \frac{1}{2} )[1- \ln( in + i\frac{1}{2} ) + (2s -1)\pi ] \}$. 
The relation $ B=B(h)$ follows from
$\varepsilon_{\sigma} (B) = 0$.
In the limit of strong coupling, 
we  also obtain a Wiener-Hopf  equation for
$\varepsilon_{\sigma}(\lambda)$ 
from the second equation of (\ref{eq:dresseqs}).
Using the similar procedure, we
solved the dependence of $B$ on $h$.  An explicit
expression for the susceptibility at zero temperature 
(\ref{eq:susceptibility}) follows with 
$\partial B /\partial h = - 1/\pi h$.

In the above we obtained exact results for the Kondo impurity
by taking into account the electronic correlations in the case of 
quadratic dispersion relation. We showed  that 
the model is integrable when the correlation strength 
$u$ is proportional to the $J$, the strengh of 
electron-impurity coupling.  If we consider
the approximation $k_j \sim k_l$ for any $j, l$, 
the $S$ matrix of electron-electron will be independent of 
$u$. Then the Yang-Baxter equation will give no relation
between $u$ and $J$. This makes it easy to understand
the usual Kondo problem where the linear dispersion
relation is adopted. 


You-Quan Li was supported by NSFC No. 19675030, 
NSFZ No. 194037, and  Y. Pao \& and Z. Pao Foundation.
The authors are grateful to C. Gruber for discussions.

\end{multicols}
\end{document}